\documentclass[useAMS,usenatbib,usegraphicx]{mn2e}
\usepackage{psfig}   
\usepackage{graphicx}
\usepackage{amssymb}
\usepackage{subfigure}

\title[The mass ratio distribution and star formation]{The binary companion mass ratio distribution: an imprint of the star formation process?}

\author[R.~J.~Parker \& M.~M.~Reggiani]{
  Richard J.~Parker\thanks{E-mail: rparker@phys.ethz.ch} and Maddalena M. Reggiani\vspace*{0.1cm}\\
   Institute for Astronomy, ETH Z{\"u}rich, Wolfgang-Pauli-Strasse 27, 8093 Z{\"u}rich, Switzerland}

\begin{document}

\date{Accepted for publication in MNRAS}
                             
\pagerange{\pageref{firstpage}--\pageref{lastpage}} \pubyear{2013}

\maketitle

\label{firstpage}

\begin{abstract}
We explore the effects of dynamical evolution in dense clusters on the companion mass ratio distribution (CMRD) of binary stars. Binary systems are 
destroyed by interactions with other stars in the cluster, lowering the total binary fraction and significantly altering the initial semi-major axis 
distribution. However, the shape of the CMRD is unaffected by dynamics; an equal number of systems with high mass ratios are destroyed compared to 
systems with low mass ratios. We might expect a weak dependence of the survivability of a binary on its mass ratio because its binding energy is proportional to 
both the primary and secondary mass components of the system. However, binaries are broken up by interactions in which the perturbing star has a significantly higher energy (by a factor of $\gtrsim 10$, depending on the particular binary properties) 
than the binding energy of the binary, or through multiple interactions in the cluster. We therefore suggest that the shape of the observed binary CMRD is an outcome of the star formation process, and should be measured in preference 
to the distributions of orbital parameters, such as the semi-major axis distribution.
\end{abstract}


\section{Introduction}

One of the major unsolved questions in astrophysics is whether the
star formation process is universal, or whether it varies as a function of the local environment. Studies of the Initial Mass Function (IMF) in nearby 
star forming regions have shown that the IMF is invariant \citep*{Bastian10}. 

A further exploration of this universality hypothesis can be undertaken by studying the properties of binary stars in different environments \citep[e.g.][]{Reggiani11,King12a,King12b}. 
If the properties of binaries vary as a function of environment, then we can conclude that star formation is not universal, and depends on some localised parameter.

However, comparing binary statistics in different star formation environments is far from straightforward. Firstly, one needs to compare the statistics of binaries over a common 
primary mass range, separation range and mass ratio sensitivity \citep[e.g.][]{King12a}. 

Secondly, dynamical evolution in dense star formation regions has been shown to alter the overall binary fraction, and the separation and orbital eccentricity distributions 
\citep*[e.g.][]{Kroupa95a,Kroupa95b,Kroupa99,Parker11c,Marks11}. It is possible to `reverse engineer' the initial conditions of star formation in clusters to determine the amount of
dynamical processing that has taken place, but in order to do this one must first make assumptions about the initial binary properties, and the highest density the cluster attained \citep{Kroupa95a,Parker09a,Marks11}. 
Such assumptions are non-trivial if the dynamical processing has occurred `locally' \citep[e.g.\,\,within pockets of substructure in the cluster,][]{Parker11c}, rather than globally.

A more fruitful approach would be to focus observational effort on a measurable parameter of binary stars that may not be affected by dynamical evolution. One such parameter could be the distribution of binary mass 
ratios, $q$, where $q = m_s/m_p$, and $m_p$ and $m_s$ are the masses of the primary and secondary components of the system, respectively. Recent work by \citet{Reggiani11,Reggiani13} compared this companion mass ratio distribution (CMRD) 
in different star formation regions, and open clusters, with binaries in the Galactic field. They find that the shape of the CMRD is consistent with being linearly flat in the field  ($dN/dq \propto q^{-0.50 \pm 0.29}$), and in most star forming regions (with the possible exception of the Taurus association). 

Dynamical interactions will reduce the overall fraction of binary systems in a cluster, but it is unclear whether the shape of the CMRD is altered. In $N$-body simulations of very low mass binaries (VLMBs) in dense clusters, 
 \citet{Parker11a} found that the CMRD for VLMBs is uniformly lowered, with no preference for destroying systems with a low $q$. However, VLMBs typically form as equal--mass systems, and it is unclear how the CMRD would 
change for binaries with higher primary masses and an initially flat CMRD, as is observed in the field.

In this paper, we investigate the effects of dynamical interactions in dense clusters on the shape of the CMRD. We summarise the analytics of binary destruction in Section~\ref{analytics}, we outline our numerical simulations in Section~\ref{method}, we present our results in Section~\ref{results}, we provide a discussion 
in Section~\ref{discuss} and we conclude in Section~\ref{conclude}.

\section{Analytics of binary destruction}
\label{analytics}

Several dynamical classes of binary exist \citep{Heggie75,Hills75a}. `Hard binaries' have a binding energy greater than the local Maxwellian energy ($\left<m\right>\sigma^2$, where $\left<m\right>$ is the average stellar mass and $\sigma$ is the Maxwellian velocity), and are unlikely to be destroyed in an interaction (indeed, they are often `hardened'). `Soft' binaries have a binding 
energy which is much lower than the local Maxwellian energy, and can be disrupted by very small perturbations \citep{Binney87}. Finally, `intermediate' binaries have a binding energy comparable to the local Maxwellian energy. The binding energy of a binary, $E_{\rm bind}$, can be written as 
\begin{equation}
E_{\rm bind} = -\frac{Gm_pm_s}{2a},
\label{binding}
\end{equation}
where $m_p$ and $m_s$ are the primary and secondary masses and $a$ is the semi-major axis. If two binaries have the 
same separation, $a$, they should have slightly different binding energies if their mass ratios are different.  If we consider two binary systems with equal separations, but one is a 
1\,M$_\odot$--1\,M$_\odot$ system and the other is a 1\,M$_\odot$--0.1\,M$_\odot$ system, then  we expect the latter system to have a lower binding energy 
(and hence may be more susceptible to destruction) than the former.

However, analytic considerations of binary destruction are further complicated when the mass ratio of the system is not unity \citep[e.g.][]{Fregeau06}. If we consider an interaction between a binary and perturbing star of mass $M_{\rm Pert}$, the critical velocity of the system (where the total energy of the binary and perturbing star is zero) can be written as:
\begin{equation}
v_c = \left(\frac{Gm_pm_s}{\mu a}\right)^{1/2},
\end{equation}
where $\mu$ is the reduced mass of the binary--perturbing star system: 
\begin{equation}
\mu = \frac{(m_p + m_s)M_{\rm Pert}}{m_p + m_s + M_{\rm Pert}}.
\end{equation}
Furthermore, if the orbital velocity, $v_{\rm orb}$ is
\begin{equation}
v_{\rm orb} = \left(\frac{G(m_p + m_s)}{a}\right)^{1/2},
\end{equation}
then for comparable masses ($m_p = m_s = M_{\rm Pert}$), $v_c \simeq v_{\rm orb}$. However, if $m_p >> m_s$, but $m_p \simeq M_{\rm Pert}$, then $v_c \simeq \sqrt{2q}v_{\rm orb}$. 

The outcome of an interaction between a single star and a binary depends on the relative velocity of the perturbing star, $v_{\rm Pert}$, the critical velocity, $v_c$, and the orbital velocity, $v_{\rm orb}$ \citep[e.g.][and references therein]{Fregeau06}. If 
$v_{\rm Pert} < v_c$, then destruction is not possible, but such a scenario favours exchange interactions (which would likely increase the mass ratio). On the other hand, if $v_c < v_{\rm Pert} < v_{\rm orb}$, then binaries can be destroyed without affecting the shape 
of the mass ratio distribution. This subtle dependence of binary destruction on both the orbital and critical velocities for systems with unequal masses can be summarised by considering the hard--soft boundary as the ``fast--slow'' boundary \citep{Hills89, Hills90}.

\section{Method}
\label{method}

\subsection{Cluster structure and virial state}
\label{morph}

We run four suites of simulations to examine the potential change in CMRD through dynamical interactions in a clustered environment. In the first three sets of simulations, 
the binary systems are placed in cool, clumpy clusters \citep*{Parker11c} because such initial conditions reflect the fact that many star forming regions are observed to be 
substructured \citep[e.g.][]{Cartwright04,Sanchez09} and subvirial \citep[e.g.][]{Peretto06,Furesz08}.  

Furthermore, \citet{Parker12b} have recently shown that these cool, clumpy clusters can stochastically alter identical binary populations so that the separation distributions are 
statistically different after 1\,Myr in different clusters. Therefore, these initial conditions represent the most extreme way of disrupting binaries and may alter the CMRD to an even greater extent than say, smooth, virialised clusters. We also run one suite of simulations with the binaries distributed in smooth, virialised Plummer spheres \citep{Plummer11} as a control run.

The substructured clusters are set up as fractals with a radius $r_F = 1$\,pc, according to the prescription detailed in \citet{Goodwin04a}. Fractals have the advantage that the amount of substructure is 
described by just one number (the fractal dimension, $D$), although it is unclear whether stars actually form in a fractal distribution \citep[they may do;][]{Elmegreen01}. Highly 
fractal clusters ($D = 1.6$) process more binaries than smoother clusters (e.g. $D = 2.6$) due to the dense pockets of substructure \citep{Parker11c}, and in this work we adopt $D = 2.0$, 
which is a `moderate' level of substructure. 

The stars in our fractal clusters have correlated velocities, so that two nearby stars in the fractal have similar velocities, whereas distant stars may have very different 
velocities, as described in \citet{Goodwin04a} and \citet{Parker11c}.

In the suite of simulations in which the stars are distributed in a smooth Plummer sphere, the positions and velocities are chosen according to the prescription in \citet*{Aarseth74}.

Finally, we scale the velocities of the stars in the cluster to the desired virial ratio ($Q_{\rm vir} = T/|\Omega|$ where $T$ and $|\Omega|$ are the total kinetic energy and total 
potential energy of the stars, respectively). For the fractal cluster simulations we choose a subvirial (cool) virial ratio, $Q_{\rm vir} = 0.3$, whereas the Plummer sphere clusters are in 
virial equilibrium ($Q_{\rm vir} = 0.5$).

\subsection{Initial binary populations}

We create the clusters so that every star is placed in a binary system initially. Primary masses are drawn from a 2--part \citet{Kroupa02} IMF of the form:
\begin{equation}
 N(M)   \propto  \left\{ \begin{array}{ll} 
 M^{-1.3} \hspace{0.4cm} m_0 < M/{\rm M_\odot}  \leq m_1   \,, \\ 
 M^{-2.3} \hspace{0.4cm} m_1 < M/{\rm M_\odot} \leq m_2   \,,
\end{array} \right.
\label{imf}
\end{equation}
where $m_0$ = 0.1\,M$_\odot$, $m_1$ = 0.5\,M$_\odot$, and  $m_2$ = 50\,M$_\odot$. We choose secondary masses according to the flat CMRD observed in the Galactic field 
\citep{Reggiani11,Reggiani13}. For a comprehensive discussion of different binary component pairing algorithms, we refer the interested reader to \citet{Kouwenhoven09a}. We do not allow brown dwarf primaries in the simulations.

In two sets of simulations we draw periods from the distribution found by \citet{Duquennoy91} for field G-dwarfs \citep[see][for an updated but similar fit]{Raghavan10}:
\begin{equation}
f\left({\rm log_{10}}P\right)  = {\rm exp}\left \{ \frac{-{({\rm log_{10}}P - \overline{{\rm log_{10}}P})}^2}{2\sigma^2_{{\rm  log_{10}}P}}\right \},
\end{equation}
where $\overline{{\rm log_{10}}P} = 4.8$, $\sigma_{{\rm log_{10}}P} =
2.3$ and $P$ is  in days. We convert the periods to semi-major axes
using the masses of the binary components. This corresponds to semi-major axes in the range $10^{-2} - 10^5$\,au.

The eccentricities of the binary orbits are drawn from a thermal
distribution \citep{Heggie75} of the form
\begin{equation}
f_e(e) = 2e.
\end{equation}
In the sample of \citet{Duquennoy91}, close binaries (with periods
less than 10 days) are almost exclusively on tidally circularised
orbits. We account for this by reselecting the eccentricity of a
system if it exceeds the following  period-dependent
value:
\begin{equation}
e_{\rm tid} = \frac{1}{2}\left[0.95 + {\rm tanh}\left(0.6\,{\rm log_{10}}P - 1.7\right)\right].
\end{equation}

In the remaining two simulations, we assign \emph{all} binaries a semi-major axis of either 30\,au or 10\,au, in order to test whether the shape of the CMRD could be altered by dynamical 
interactions when the semi-major axis becomes a constant in Equation~\ref{binding}.

We combine the primary and secondary masses of the binaries with their
semi-major axes and eccentricities to determine the relative velocity
and radial components of the stars in  each system. The binaries are
then placed at the centre of mass and velocity for each system in
either the fractal distribution or Plummer sphere (see
Section~\ref{morph}). 

The simulations are run for 10\,Myr  using the \texttt{kira} integrator in the Starlab package \citep[e.g.][]{Zwart99,Zwart01}. We do not include stellar evolution in the
simulations. A summary of the four sets of simulations is given in Table~\ref{simulations}.

\begin{table}
\caption[bf]{A summary of the four sets of simulations.
The values in the columns are: the number of stars in each cluster ($N_{\rm stars}$), the morphology of the cluster (either a Plummer sphere or fractal), the initial virial ratio of the cluster ($Q_{\rm vir}$), the initial radius of the fractal, ($r_{\rm F}$), or the initial half-mass radius of the Plummer sphere, ($r_{1/2}$), the fractal dimension, $D$ (if applicable) and the 
adopted semi-major axis distribution, $a_{\rm bin}$ (either \citet{Duquennoy91},  a delta function at 30\,au or a delta function at 10\,au).}
\begin{center}
\begin{tabular}{|c|c|c|c|c|c|}
\hline 
$N_{\rm stars}$  & Morphology & $Q_{\rm vir}$ &  $r_{\rm F}$ or $r_{1/2}$ & $D$ & $a_{\rm bin}$ \\
\hline
1500 & Fractal & 0.3 & 1\,pc & 2.0 & DM91 \\
1500 & Fractal & 0.3 & 1\,pc & 2.0 & 30\,au \\
1500 & Fractal & 0.3 &  1\,pc & 2.0 & 10\,au \\
\hline
1500 & Plummer sphere & 0.5 & 0.1\,pc & -- & DM91 \\
\hline
\end{tabular}
\end{center}
\label{simulations}
\end{table}

\section{Results}
\label{results}

In this section we first compare the shape of the semi-major axis distribution and companion mass ratio distribution (CMRD) at 0 and 10\,Myr (i.e.\,\,before and after dynamical evolution in the cluster). We then consider the interaction histories of binaries that are destroyed.

\subsection{Initial and final distributions}

We compare the initial (0\,Myr) semi-major axis (hereafter separation) distribution and CMRD to the distributions after 10\,Myr. Most binary destruction in the type of clusters we simulate here occurs in the first 1\,Myr \citep[e.g.][]{Parker09a,Marks11,Parker11c}, but we analyse the simulations at 10\,Myr to ensure that all destructive encounters are tracked. After 10\,Myr the clusters are so diffuse that very little further dynamical processing will occur before the binaries become members of the Galactic field population.  We determine whether a binary is energetically bound or not using the 
nearest neighbour algorithm outlined in \citet{Parker09a} and \citet{Kouwenhoven10}. The number of binaries, and the binary fractions at 0 and 10\,Myr are given in Table~\ref{bin_results} for each simulation.

\begin{figure*}
  \begin{center}
\setlength{\subfigcapskip}{10pt}
\hspace*{-0.3cm}
\subfigure[Separation distribution]{\label{field_bin_sep}\rotatebox{270}{\includegraphics[scale=0.37]{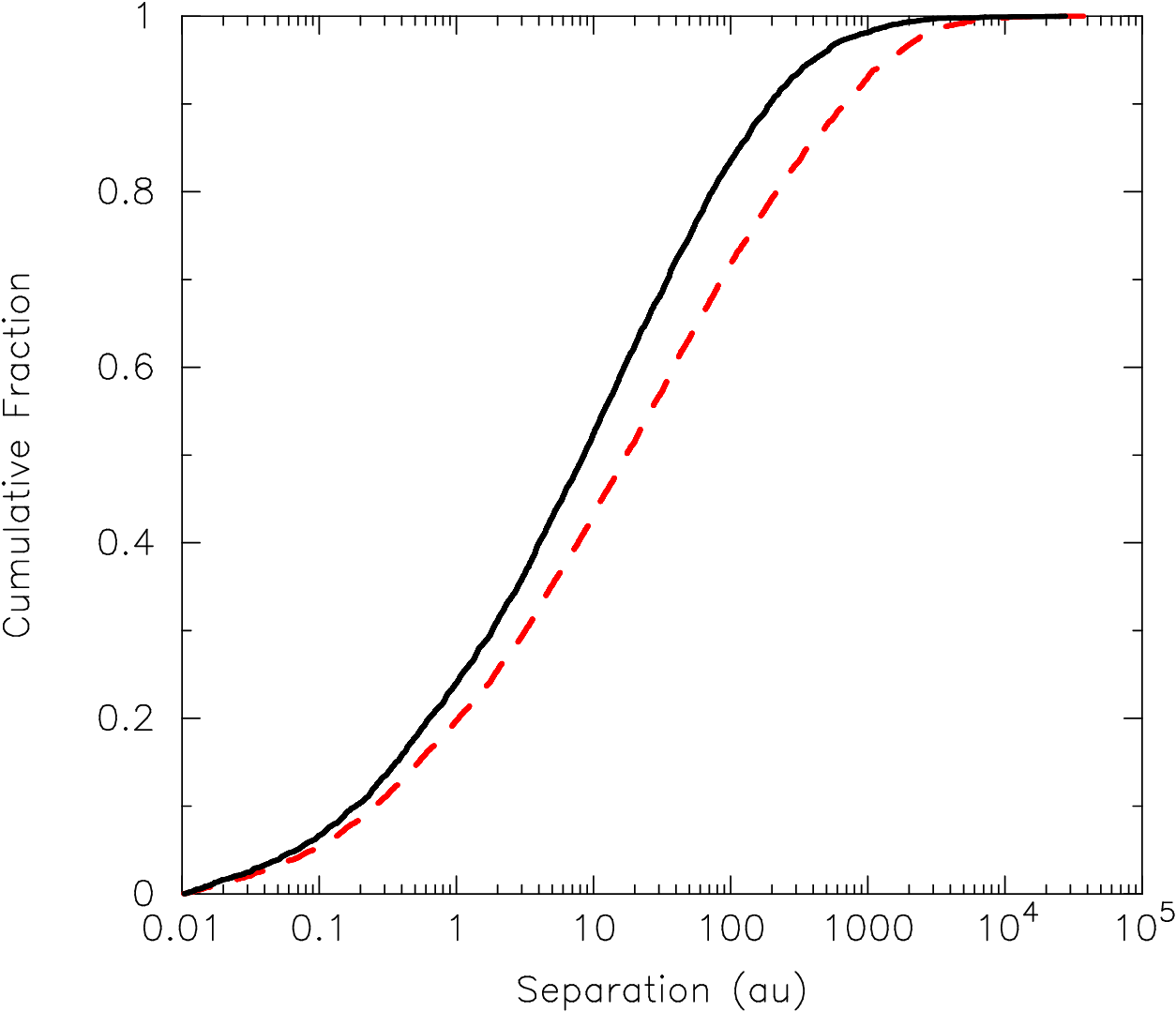}}}  
\hspace*{0.2cm}
\subfigure[Mass ratio distribution]{\label{field_bin_CMRD}\rotatebox{270}{\includegraphics[scale=0.37]{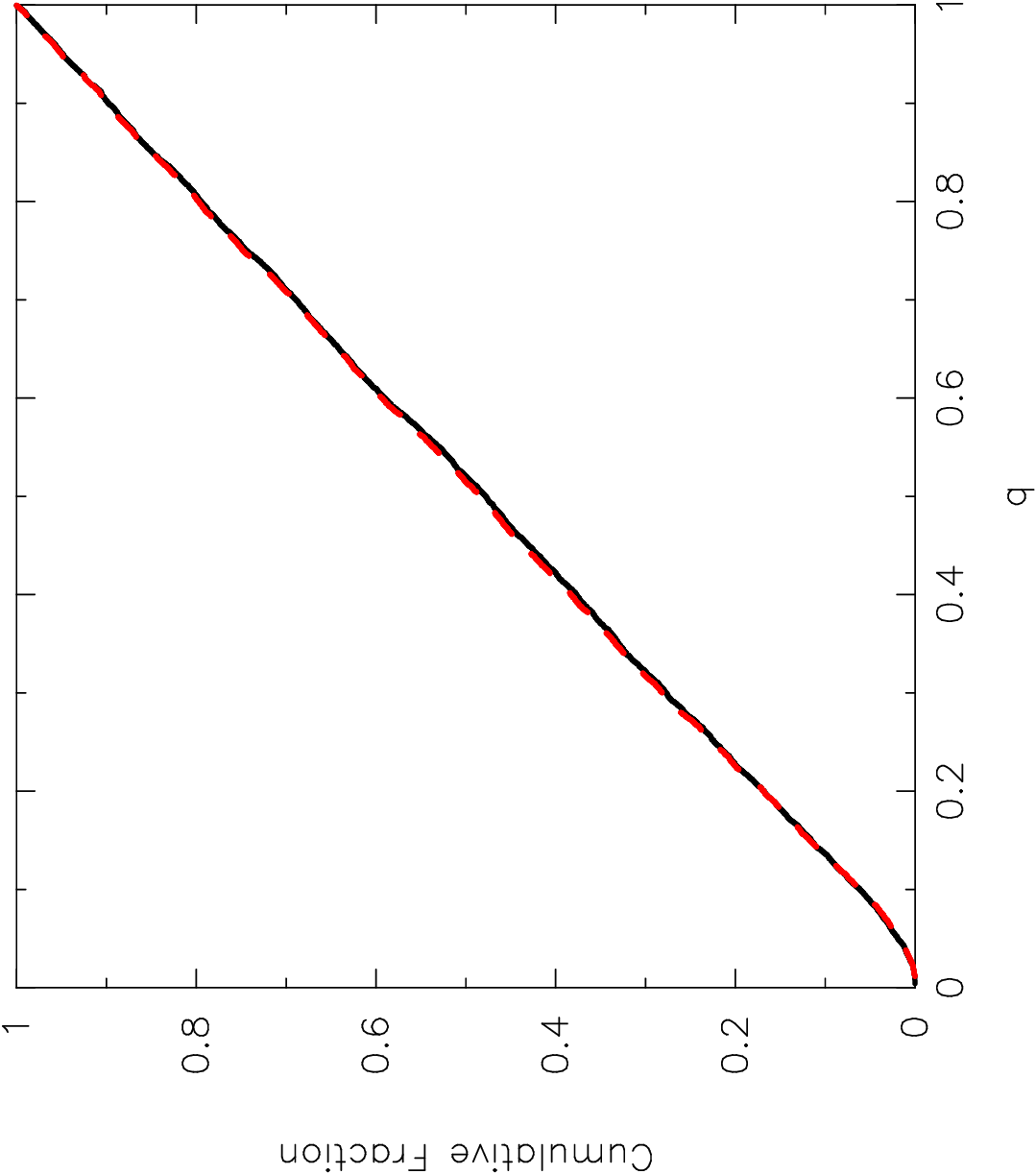}}}
\end{center}
  \caption[bf]{Evolution of the binary cumulative separation distribution and cumulative companion mass ratio distribution (CMRD) for binaries with periods drawn from 
the \citet{Duquennoy91} fit to the Galactic field data. The initial distributions at 0\,Myr are shown by the red dashed lines, and the distributions after 10\,Myr of dynamical 
evolution are shown by the solid black lines.}
  \label{field_bin_results}
\end{figure*}

In our first simulation, we placed 750 binary systems with a field-like separation distribution in a cool, clumpy star cluster. Such clusters violently relax during their collapse 
to a centrally concentrated, spherical cluster \citep[see figs.~1~and~2 in][]{Parker11c}, and binaries are processed during the erasure of substructure and due to the dense core 
which forms. In Fig.~\ref{field_bin_results} we show the evolution of (a) the separation distribution and (b) the CMRD. The initial distributions are shown by the red dashed lines 
and the final (10\,Myr) distributions are shown by the solid black lines.

\begin{table*}
\caption[bf]{A summary of the numbers of binaries destroyed in each simulation. The values in the columns are: the simulation number, the morphology of the cluster (either a Plummer sphere or fractal), the initial virial ratio of the cluster ($Q_{\rm vir}$), 
the adopted semi-major axis distribution, $a_{\rm bin}$ (either \citet{Duquennoy91}, a delta function at 30\,au or a delta function at 10\,au), the number of binaries at 0\,Myr ($N_{\rm bin, 0\,Myr}$), the binary fraction at 0\,Myr ($f_{\rm bin, 0\,Myr}$), the 
number of binaries at 10\,Myr ($N_{\rm bin, 10\,Myr}$) and the binary fraction at 10\,Myr ($f_{\rm bin, 10\,Myr}$).}
\begin{center}
\begin{tabular}{|c|c|c|c|c|c|c|c|}
\hline 
Sim.\,\,No.  & Morphology & $Q_{\rm vir}$ & $a_{\rm bin}$ & $N_{\rm bin, 0\,Myr}$ & $f_{\rm bin, 0\,Myr}$ & $N_{\rm bin, 10\,Myr}$ & $f_{\rm bin, 10\,Myr}$\\
\hline
1 & Fractal & 0.3 & DM91 & 686 & 0.85 & 534 & 0.60 \\
2 & Fractal & 0.3 & 30\,au & 750 & 1.00 & 647 & 0.80 \\
3 & Fractal & 0.3 & 10\,au & 750 & 1.00 & 700 & 0.90 \\
\hline
4 & Plummer sphere & 0.5 & DM91 & 654 & 0.78 & 512 & 0.53 \\
\hline
\end{tabular}
\end{center}
\label{bin_results}
\end{table*}

Clearly, the shape of the separation distribution changes during the first 10\,Myr of evolution. Binaries are destroyed (the binary fraction decreases from  a primordial value of 85\,per cent\footnote{Note that formally we placed every star in a binary initially. However, the initial cluster density causes the widest binaries to be unbound, even before dynamical evolution. These initially unbound binaries are not included in the analysis.} to 60\,per cent after 10\,Myr), especially the wider systems ($\gtrsim$ 200\,au) 
which are intermediate--soft in such dense environments according to the Heggie--Hills law \citep{Heggie75,Hills75a} and are broken up. However, the CMRD retains the same 
shape, even after dynamical evolution. We find (almost) identical results in the simulations of Plummer-sphere clusters (comparison of the evolution of the binary fraction and separation distribution for Plummer sphere clusters \citep{Parker09a} and for fractal clusters \citep{Parker11c} also shows that the amount of dynamical evolution in both morphologies is similar).

\begin{figure*}
  \begin{center}
\setlength{\subfigcapskip}{10pt}
\hspace*{-0.3cm}
\subfigure[Separation distribution]{\label{delta_bin_sep}\rotatebox{270}{\includegraphics[scale=0.37]{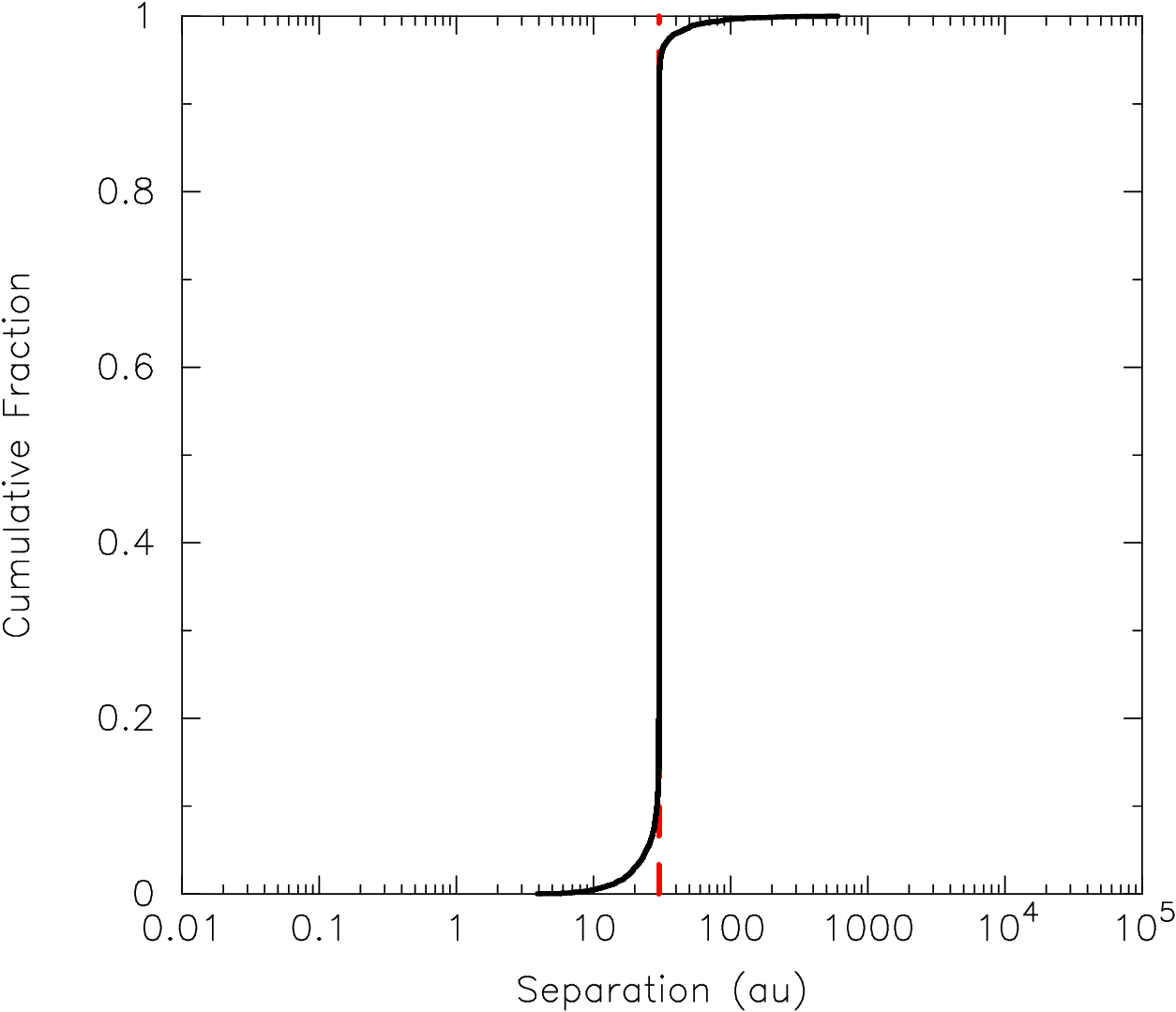}}}  
\hspace*{0.2cm}
\subfigure[Mass ratio distribution]{\label{delta_bin_CMRD}\rotatebox{270}{\includegraphics[scale=0.37]{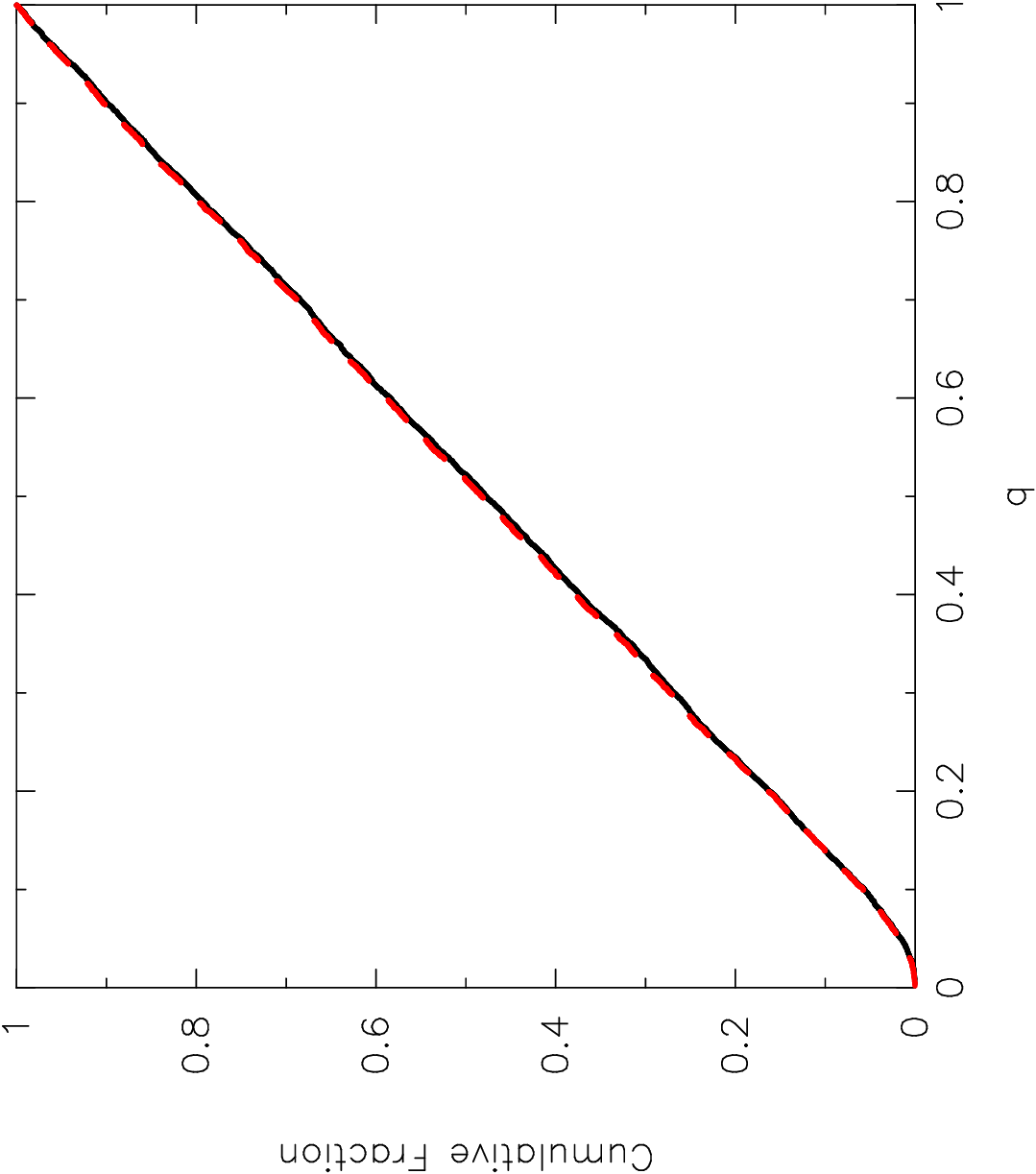}}}
\end{center}
  \caption[bf]{Evolution of the binary cumulative separation distribution and cumulative companion mass ratio distribution (CMRD) for binaries with initial separations of 30\,au. 
The initial distributions at 0\,Myr are shown by the red dashed lines, and the distributions after 10\,Myr of dynamical 
evolution are shown by the solid black lines.}
  \label{delta_bin_results}
\end{figure*}

In Fig.~\ref{delta_bin_results} we investigate whether placing binaries in clusters with identical (30\,au) separations results in different initial and final CMRDs. Again, a 
certain number of binaries are broken up (the binary fraction decreases from 100\,per cent to  80\,per cent after 10\,Myr), and a smaller fraction have their separation hardened or softened (see panel (a) of  Fig.~\ref{delta_bin_results}), but the shape 
of the CMRD is unchanged (Fig.~\ref{delta_bin_CMRD}).

\subsection{Binary disruption energetics}

These results indicate that destruction of binaries is independent of the mass ratio, $q$. However, according to Equation~\ref{binding}, if two binaries have the 
same separation, they should have slightly different binding energies if their mass ratios are different.  

\begin{figure}
\begin{center}
\rotatebox{270}{\includegraphics[scale=0.3]{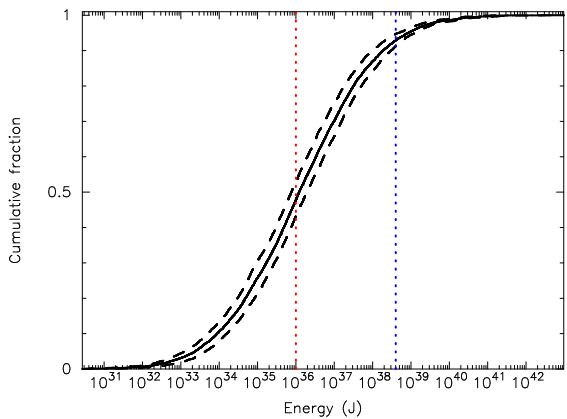}}
\end{center}
\caption[bf]{Cumulative distribution of binary binding energies for systems with primary masses drawn from a \citet{Kroupa02} IMF, secondary masses drawn from a flat mass ratio distribution \citep{Reggiani11} and periods from the \citet{Duquennoy91} distribution observed in the Galactic field. We show the distribution for 
all systems by the solid line, systems with $q \leq 0.5$ by the left-hand dashed line, and systems with $q > 0.5$ by the right-hand dashed line (as binding energy weakly depends on the mass ratio). We show the kinetic energy of a 1\,M$_\odot$ star travelling at (a) 1\,km\,s$^{-1}$ with the lefthand dotted red line and (b) 
20\,km\,s$^{-1}$ with the right hand dotted blue line. }
\label{bin_energy}
\end{figure}

We demonstrate this in Fig.~\ref{bin_energy}, where we draw binaries randomly from the \citet{Duquennoy91} semi-major axis distribution, choosing the primary from the 
IMF and assigning the secondary from a flat companion mass ratio distribution \citep{Reggiani11}. The solid line shows the cumulative distribution of binding energies for all 
binaries, and the dashed lines show the distributions for systems with $q \leq 0.5$ (the left-hand line) and $q > 0.5$ (the right hand line). 

The binding energy is weakly dependent on $q$, but of far more importance is the energy of a star which may break up the binary. 
A binary tends to be broken up in an encounter if the kinetic energy of the perturber exceeds the binding energy of the binary. We show the kinetic energy of a 1\,M$_\odot$ star travelling at (a) 1\,km\,s$^{-1}$ (the lefthand dotted red line) and (b) 20\,km\,s$^{-1}$ (the right hand dotted blue line). Stars in clusters typically have 
velocities $\sim$1\,km\,s$^{-1}$, but can suffer dynamical interactions which significantly increase their kinetic energy. Fig.~\ref{bin_energy} shows that the differences in binding energy due to low mass ratios are small compared to the range of kinetic energies a star can have in a clustered environment.

 With this in mind, we examine the distribution of encounter energies that break up binaries. We track the interaction history of every binary in our simulations, and then determine the kinetic energy, $E_{\rm pert}$, of the interaction which destroys the binary:
\begin{equation}
E_{\rm pert} = \frac{1}{2}M_{\rm pert}v_{\rm pert}^2.
\end{equation}
Here, $M_{\rm pert}$ is the mass of the perturbing star, and $v_{\rm pert}$ is the magnitude of the relative velocity of the perturbing star with respect to the binary. If the perturbing star happens to also be a binary, we calculate $v_{\rm pert}$ using the centre of mass velocity of that binary, and use the mass of the binary for $M_{\rm pert}$.

\begin{figure}
\begin{center}
\rotatebox{270}{\includegraphics[scale=0.35]{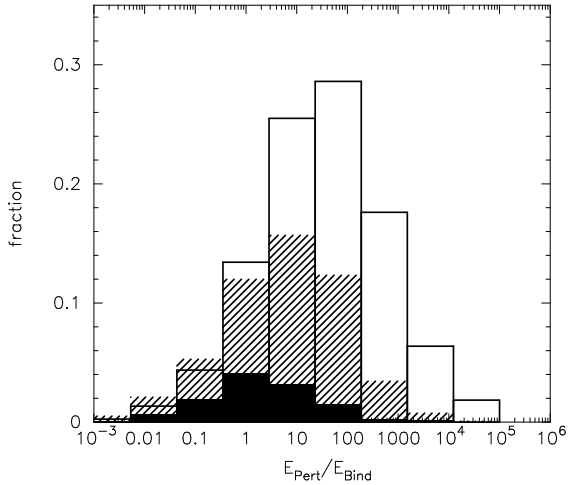}}
\end{center}
\caption[bf]{Distribution of the kinetic energy of the interaction between a perturbing star and a binary which destroys the binary, divided by the binding energy of the binary, for three of our simulations. The open histrogram is the distribution for the simulation which has binary separations drawn from \citet{Duquennoy91}, the hashed histogram is for the simulation which has all binaries with 30\,au separations and the solid histogram is the simulation where all binaries have 10\,au separations. The histograms are normalised to the number of binaries that are destroyed after 10\,Myr in the first simulation.}
\label{energy_hist}
\end{figure}

In Fig.~\ref{energy_hist}, we show the distribution of the kinetic energy of the interaction between a perturbing star and a binary which destroys the binary, divided by the binding energy of the binary. The open histrogram is the distribution for the simulation which has binary separations drawn from \citet{Duquennoy91}, the hashed histogram is for the simulation which has all binaries with 30\,au separations and the solid histogram is the simulation where all binaries have 10\,au separations. The histograms are normalised to the number of binaries that are destroyed between 0 and 10\,Myr in the first simulation (with the field-like separation distribution) to demonstrate the decrease in the number of systems that are destroyed with decreasing binary separation. 

For the simulation with separations drawn from \citet{Duquennoy91}, the distribution of perturbing kinetic energy to binding energy peaks at a ratio $E_{\rm pert}/E_{\rm bind} \sim 10 - 100$, suggesting that the input energy required to destroy a binary typically exceeds the binding energy by a factor similar to, or in excess of the most extreme mass ratios. However, the distributions for the simulations with separations drawn from delta functions of 30\,au and 10\,au (the hashed and solid histograms, respectively) indicate that for lower separations, the ratio of $E_{\rm pert}/E_{\rm bind}$ decreases towards unity. This could in principle lead to evolution of the CMRD, as the peak of the distribution of $E_{\rm pert}/E_{\rm bind}$  moves to values similar to the difference in binding energy between systems with vastly different mass ratios. In practice, this is unlikely; so few binaries are destroyed in the cluster with binary separations at 10\,au (typically only 50 out of an initial total of 750) that a drastic bias towards destroying systems with low $q$ would be required before this became evident in the distribution.

A small fraction (usually less than 10\,per cent, depending on the initial separation distribution) of the destroyed binaries have an $E_{\rm pert}/E_{\rm bind}$ ratio less than unity. These binaries are destroyed by multiple interactions, which have the progressive effect of lowering the binding energy and leaving the binary more susceptible to destruction after each encounter.

Finally, we note that the shape of the CMRD could be altered by exchange interactions (for example, if $v_{\rm Pert} < v_c$). In this scenario, the least massive star in the binary is replaced by a more massive intruder, and the mass ratio will therefore increase. In our simulations, we find that typically $\lesssim 10$ exchange interactions occur per cluster, which is too small a rate to affect the shape of the CMRD. 

\section{Discussion}
\label{discuss} 

Using $N$-body simulations, we follow the dynamical 
evolution of clusters containing 750 primordial binary stars for 10\,Myr. In our simulations of clusters, which are initially subvirial and substructured, binaries are destroyed immediately on local scales (i.e. within the substructure), before 
the global collapse of the cluster at around $\sim$0.8\,Myr \citep[e.g.][]{Allison10,Parker11c}. The results presented in Section~\ref{results} indicate that the shape of the binary star companion mass ratio distribution (CMRD) is unaffected by dynamical evolution in dense clusters.

The (lack of) evolution of the CMRD is also unaffected by the assumed initial orbital parameters of binary systems. In the simulations presented in Fig.~\ref{delta_bin_results} we created binaries with separations drawn from a delta function at 30\,au. Although we would 
expect the binding energy of binary systems to be weakly dependent on the binary mass ratio, in practice the shape of the CMRD remains constant. The reasons for this are twofold. Firstly, the average energy of perturbing stars (travelling on average at $\sim$1\,km\,s$^{-1}$) in the 
clustered environment is higher than the binding energy of half of the binaries in the cluster, with only a weak dependence on mass ratio. Secondly, the interactions that do destroy binary systems typically have a perturbing energy of order 10 -- 100 times the binding energy; i.e. they are so 
energetic that the dependence of the binding energy on mass ratio is irrelevant. Finally, a binary may suffer multiple interactions in a clustered environment, and can be de-stablised by previous interactions, before the encounter that finally destroys it.

As we have shown in Fig.~\ref{energy_hist}, the harder a binary, the lower the ratio of perturbing energy to binding energy is. Therefore, there may be a binary separation regime ($<$\,10\,au in the clusters we simulate here) in which the CMRD could change. However, at such low separations very few binaries are destroyed, and so the shape of the CMRD would be expected to remain roughly constant. A similar paucity of exchange interactions in these clusters also prevents the shape of the CMRD changing towards larger $q$ values.

Previous work on dynamical evolution of binaries in star clusters suggested that the shape of the CMRD \emph{is} altered by dynamics. \citet{Kroupa95a} plotted the initial and final distributions of the secondary mass ($m_s$) component of binaries with a G-type primary ($0.8 < m_p/{\rm M}_\odot < 1.2$) and showed 
that more binaries with $m_s \sim 0.2$\,M$_\odot$ are destroyed overall (see his fig.~4). However, \citet{Kroupa95a} chose secondary masses from randomly sampling an IMF, the peak of which lies around $\left<m\right> = 0.2$\,M$_\odot$. Therefore, it is unsurprising that most binaries that are destroyed have a relatively low-mass companion and 
such a plot does not show the evolution of the mass ratio distribution. When one compares the \emph{shape} of the CMRD, however, it is clear that dynamics does not alter an initial distribution.

More recent work by \citet{Fregeau04} examined binary--binary and binary--single star interactions in globular clusters, and found that destruction is only very weakly dependent on the 
binary mass ratios\footnote{A wealth of literature also exists which presents numerical scattering experiments examining the effects of binary--binary and binary--single star interactions \citep[e.g.][]{Hut83b,Mikkola83,Hills90,Heggie93}. However, these experiments tend to focus on systems with equal mass ($m_p = m_s$; $q = 1$) binary components, though see \citet*{Sigurdsson93,Heggie96b} for a description of exchange interactions with unequal mass components. This, combined with a lack of information on subsequent encounters (which may further de-stablise soft binaries and could be important in a clustered environments) makes it difficult to compare the results to $N$-body simulations of binary destruction in clusters.}. In direct $N$-body models of the old open cluster NGC\,188, \citet*{Geller13} also found the initial and final CMRDs for main sequence binaries to be statistically indistinguishable.

It should also be noted that the clustered environments modelled here have high densities (by design, so that some binaries are destroyed and we can examine the change in the CMRD) compared to those in the local Solar neighbourhood \citep{Bressert10}, where the Orion Nebula Cluster (ONC) is the most dense region. 
Based on the \citet{Bressert10} data, \citet{Parker12d} recently estimated that up to 50\,per cent of star forming events could be dense enough to affect binary systems, so it is possible that many binary systems do not undergo significant dynamical processing. We have shown here that even in clusters with similar densities to the ONC (extrema in terms of local star formation), dynamical interactions are unlikely to affect the shape of the CMRD.

Furthermore, we have not applied any cuts to our simulations in terms of primary mass, separation or mass ratio ranges, which are often applied to observational samples to make them consistent \citep{Reggiani11}. Typically, observations of visual binaries in clusters span the range 10s -- 100s\,au \citep{King12a,King12b}, and we have already seen that the shape of the CMRD is not affected by dynamical evolution in this separation range. If we were to impose 
a primary mass range cut, this would not affect our results as such a cut would bias our sampled $q$ toward unity, thereby negating the argument of systems with lower binding energies being more susceptible to destruction.

There is observational evidence for a universal CMRD over a wide range of primary masses and $q$ values \citep{Metchev09}. This appears to be true both in the field and in some associations and star forming regions \citep{Reggiani11,Reggiani13}.
However, due to its independence of dynamical evolution, the CMRD may help trace differences between star formation events and test models of binary formation. If differences do exist, the combined study of the CMRD in different environments can be also used to determine which types of clustered configuration contribute most to the field population \citep[see also][]{Goodwin13}.

Finally, we note that uncovering evidence for a change in the shape of the field CMRD towards random pairing for wide binaries would support the hypothesis of  wide binary formation during the dissolution phase of star clusters \citep{Kouwenhoven10,Moeckel10,Moeckel11}, as such systems effectively form via capture rather than core or 
disc fragmentation.

\section{Conclusions}
\label{conclude}

We have conducted $N$-body simulations in which we place a population of 750 primordial binaries in a star cluster to examine the effects of dynamical interactions on the shape of the binary companion mass ratio distribution (CMRD). 
Our conclusions are the following:\\

(i) Whilst the overall fraction of binaries decreases due to destructive encounters, the shape of CMRD does not change and is independent of dynamical evolution.

(ii) We might expect that systems with similar separations but lower mass ratios would have a lower binding energy than systems with higher mass ratios, and hence be more susceptible to destruction. This is not a strong effect, however, and binaries are typically destroyed by a perturber with kinetic energy well in excess of the binding energy.

(iii) Any differences in the observed CMRD between different star forming regions would indicate different modes of star formation, and hence whether star formation is universal or not. 
The CMRD is therefore a stronger diagnostic for searching for different modes of star formation than using e.g.\,\,the binary separation distribution, because the shape of the separation distribution does change through dynamical interactions.

\section*{Acknowledgements}

We thank the anonymous referee for their comments and suggestions, which have greatly improved the paper. We also thank Michiel Cottaar and Michael Meyer for helpful discussions. The simulations in this work were performed on the \texttt{BRUTUS} computing cluster at ETH Z{\"u}rich.

\bibliographystyle{mn2e}
\bibliography{general_ref}

\label{lastpage}

\end{document}